\documentclass[twocolumn,showpacs,showkeys,preprintnumbers,amsmath,amssymb,prl]{revtex4}

\usepackage{graphicx}
\usepackage{dcolumn}
\usepackage{bm}
\tighten

\begin{document}


\title{Nonassociativity, Dirac monopoles and Aharonov-Bohm effect}

\author{A. I. Nesterov}
   \email{nesterov@cencar.udg.mx}
\affiliation{Departamento de F{\'\i}sica, CUCEI, Universidad de
Guadalajara, Av. Revoluci\'on 1500, Guadalajara, CP 44420, Jalisco,
M\'exico}

\date{\today}

\begin{abstract}
The Aharonov-Bohm (AB) effect for the singular string associated with the Dirac
monopole carrying an arbitrary magnetic charge is studied. It is shown that the
emerging difficulties in explanation of the AB effect may be removed by
introducing nonassociative path-dependent wave functions. This provides the
absence of AB effect for  the Dirac string of nonquantized magnetic monopole.

\end{abstract}



\maketitle

Nowadays interest in the Dirac monopole problem \cite{Dir} is growing in
connection with the `fictitious' monopoles that are similar to the `real'
magnetic monopoles, however, appearing in the context of the Berry phase
\cite{Berry}. Recently, the experimental results providing evidence for the
`fictitious' magnetic monopole in the crystal-momentum space has been reported
in relation to the anomalous Hall effect \cite{FNT}. Besides of the anomalous
Hall effect, these type of magnetic monopoles emerges in trapped $\Lambda$-type
atoms, anisotropic spin systems, noncommutative quantum mechanics,  etc., and
{\em may carry an arbitrary `magnetic' charge
}\cite{Br,ZLS,Hal,FP,SR,MSN,BM,FNT}.

It is known that the Dirac quantization condition $$2\mu =n, \; n \in \mathbb
Z,$$ where $\mu =eq$, electric charge  being $e$ and magnetic charge $q$ (we
set $\hbar = c=1$), does not follow from the quantum-mechanical consideration
alone. Any treatment uses some additional assumptions that may be not
physically inevitable. For instance, in Dirac's theory `quantization of
magnetic charge' follows from the requirement of the wave function be
single-valued. However, single-valudness is not one of the fundamental
principles of quantum mechanics, and having multi-valued wave functions may be
allowed until it does not affect the algebra of observables.

Other well-known topological and geometrical arguments in behalf of Dirac
quantization rule is based on employing classical fibre bundle theory
\cite{Wu1,Wu2,Gr,Gr1,G1,G2,G3}. In this approach the Dirac monopole is treated
as the Hopf bundle $U(1)$ over $S^2$.  A realization of the Dirac monopole in
this way implies that there exists the division of space into overlapping
regions $\{U_i\}$ with nonsingular vector potential being defined in $\{U_i\}$
and yielding the correct monopole magnetic field in each region. On the triple
overlap $U_i\cap U_j\cap U_k$ it holds
\begin{equation}
\exp(i(q_{ij}+q_{jk}+q_{ki})) =\exp(i4\pi\mu)
\end{equation}
where $q_{ij}$ are the transition functions such that $U_i\cap U_j\rightarrow
U(1)$. The consistency condition, which is equivalent to the associativity of
the group multiplication, requires $q_{ij}+q_{jk}+q_{ki} = 0 \mod2\pi \mathbb
Z$. This gives rise to the Dirac selectional rule $2\mu \in \mathbb Z$ as a
necessary condition to have a consistent $U(1)$-bundle over $S^2$
\cite{Wu1,Wu2,Gr}.

Finally, group-theoretical approach requires Dirac's quantization as a
necessary  condition that the operator of total angular momentum generates a
finite-dimensional representation of the rotation group
\cite{Ch,Gol1,Gol2,Zw_1,Zw_2,H}.

At the first sight these results restrict the magnetic monopole's world by the
monopoles which just satisfy the Dirac quantization condition. Nevertheless,
the self-consistent theory of Dirac's monopole with an arbitrary magnetic
charge can be constructed  nonassociative structures such as quasigroups and
loops \cite{N1,N1a,N2,NF}. To this end one needs to consider the {\em
nonassociative} generalization of $U(1)$ bundle over $S^2$ employing
nonassociative fibre bundle theory \cite{N1,N2,NF,N1a}, and in the context of
the group theory one has to consider {\em infinite-dimensional representations}
of the rotation group \cite{NF1,NF,NF3}

As is well known, any choice of the vector potential yielding a magnetic field
of the Dirac monopole must have singularity, known as the Dirac string. In this
the exploration of the Aharonov-Bohm \cite{AB} (AB) effect for the Dirac
singular string has been of great interest. The AB effect is appeared in
quantum interference between two parts of a beam of charged particles, say
electrons with charge $e$, passing by an infinite long solenoid. The
interference pattern on the screen does not change if the relative phase shift
$\Delta \varphi = e\Phi$, where $\Phi$ is the total magnetic flux through
solenoid, satisfies the following condition: $\Delta \varphi = 2 \pi n, \, n\in
\mathbb Z$.

What makes difference between the infinite long solenoid and Dirac's string is
that the latter can be moved out of position by a singular gauge
transformation. This means that the monopole string can not be observed in AB
experiment (if singular gauge transformations are allowed). Thus, the absence
of AB effect for the Dirac string has a crucial significance for a consistent
magnetic monopole theory, and, as it is well-known, this requirement yields the
Dirac quantization condition.

This raises the question of whether  the absence of AB effect for the Dirac
string is compatible with arbitrariness of magnetic monopole charge. In this
Letter we show that the response is affirmative, but it requires employment of
nonassociative structures like quasigroups and loops. \\

{\em Magnetic monopole preliminaries.} -- Since any choice of the vector
potential $\mathbf A$ being compatible with a magnetic field ${\mathbf B}= q
{\mathbf r}/r^3$ of Dirac monopole must have the singular string, one has
$\mathbf B = \nabla \times \mathbf A$ locally, but not globally, and this
implies
\begin{equation}{\mathbf
B}={\rm rot}{\mathbf A} + {\mathbf h} \label{A_4}
\end{equation}
where ${\mathbf h}$ is the magnetic field of the Dirac string $\mathcal C$ .

There is an ambiguity in the definition of the vector potential. For instance,
Dirac introduced the vector potential as \cite{Dir}
\begin{equation}
{\mathbf A}_{\mathbf n}= q\frac{{\mathbf r}\times {\mathbf n}} {r(r - {\mathbf
n} \cdot{\mathbf r})} \label{d_str}
\end{equation}
where the unit vector $\mathbf n$ determines the direction of string (hereafter
denoted by $S_{\mathbf n}$), which passes from the origin of coordinates to
$\infty$ , and
\begin{equation}
{\mathbf h}_{\mathbf n}  = 4\pi q{\mathbf n}\int _{0}^\infty \delta^3(\mathbf r
- \mathbf n \tau) d \tau.
\end{equation}
Schwinger's choice is \cite{Sw_1}:
\begin{equation}
{\mathbf A^{SW}}= \frac{1}{2}\bigl({\mathbf A}_{\mathbf n}+ {\mathbf
A}_{-\mathbf n} \bigr), \label{sw}
\end{equation}
with the string being propagated from $-\infty$ to $\infty$. Both vector
potentials yield the same magnetic monopole field, however the quantization is
different. The Dirac condition is $2\mu=p$, while the Schwinger one is $\mu=p,
\; p\in \mathbb Z$.

These two strings belong a family $\{{S}^{\kappa}_{\mathbf n} \}$ of {\it
weighted strings}, which magnetic field is given by \footnote{Previously
\cite{NF,NF1}, we have used the other definition of the string weight, namely,
${\mathbf h}^{\kappa}_{\mathbf n}= \kappa{\mathbf h}_{\mathbf n} + (1-
\kappa){\mathbf h}_{-\mathbf n}$. To compare them, one should make substitution
$\kappa \rightarrow (1+\kappa)/2$.}
\begin{align}
&{\mathbf h}^{\kappa}_{\mathbf n}= \frac{1+ \kappa}{2}{\mathbf h}_{\mathbf n} +
\frac{1- \kappa}{2}{\mathbf h}_{-\mathbf n} \label{str}
\end{align}
where $\kappa$ is the weight of a semi-infinite Dirac string. The respective
vector potential reads
\begin{align}
{\mathbf A}^\kappa_{\mathbf n} = \frac{1+ \kappa}{2}{\mathbf A}_{\mathbf n} +
\frac{1- \kappa}{2}{\mathbf A}_{-\mathbf n}. \label{A_1}
\end{align}
Note that two arbitrary strings $S^\kappa_{\mathbf n}$ and $S^\kappa_{\mathbf
n'}$ are related by gauge transformation:
\begin{eqnarray}
A^{\kappa'}_{\mathbf n'}= A^\kappa_{\mathbf n}+ d\chi. \label{ag2a},
\end{eqnarray}
and the corresponding wave function transforms as follows: $\psi' =
\exp(-i\chi)\psi$. In fact, in Dirac's approach and subsequent development of
the monopole theory the wave functions are considered as having a
non-integrable (or path-dependent) phase factor \cite{Dir,M,Wu1,Wu2}\\

\begin{figure}[bth]
\begin{center}
\scalebox{0.7}{\includegraphics{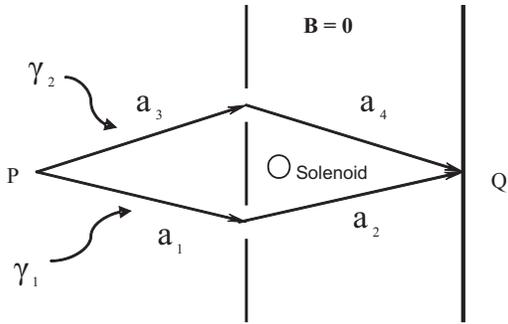}}
\end{center}
\caption{The simplified scheme of the Aharonov-Bohm experiment. The magnetic
field $\mathbf B =0$ outside the solenoid.} \label{solenoid}
\end{figure}

{\em Aharonov-Bohm effect, Dirac's string and nonassociative algebra of
observables.} -- Let a coherent beam of electrons with charge $e$ is incoming
from $-\infty$. The beam is split at the point $P$ in two parts, passing by an
infinite long solenoid (Fig.\ref{solenoid}). In spite of the fact that the
magnetic field $\mathbf B $ outside the solenoid is equal to zero, it produces
an interference effect at the point $Q$ of the screen, where  the beams are
brought together \cite{AB}.

Following Mandelstam \cite{M}, let us consider the path-dependent wave function
\begin{align}
\Psi (\mathbf r, \gamma) =  e^{i\alpha_1(\mathbf r, \gamma)}\psi(\mathbf r),
\label{eq12}
\end{align}
where $\psi(\mathbf r)$ denotes the wave function in the absence of the
magnetic field, $\mathbf A = 0$, and $\alpha_1(\mathbf r, \gamma)=e
\int_{\mathbf r}^{\mathbf r'} \mathbf A \cdot d \mathbf r $, the integration is
performed along the path $\gamma$ connecting $\mathbf r$ and $\mathbf r'$:
$\mathbf r \stackrel{\gamma}{\rightarrow}\mathbf r'$.

The total wave function at the point $Q$ of the screen is the superposition of
the wave functions along the both paths
\begin{align}
\Psi_Q& = \Psi_1(\mathbf r, \gamma_1) + \Psi_{2}(\mathbf r, \gamma_2) \nonumber\\
&= e^{i\alpha_1(\mathbf r, \gamma_1)}\psi_1(\mathbf r)+  e^{i\alpha_1(\mathbf
r, \gamma_2)}\psi_{2}(\mathbf r) \nonumber \\
&=  e^{i\alpha_1(\mathbf r, \gamma_2)}\big( e^{ie\oint \mathbf A \cdot d
\mathbf r }\psi_1(\mathbf r) +\psi_{2}(\mathbf r)\big ), \label{AB}
\end{align}
where $\alpha_1(\mathbf r, \gamma_1)$ and $\alpha_1(\mathbf r, \gamma_2)$ are
equal to $-e \int_P^Q \mathbf A \cdot d \mathbf r $  along the paths of the
first and second beam respectively.

A relative phase shift $ \Delta \varphi$ is given by
\begin{equation}
\Delta \varphi = e\oint_\gamma\mathbf A \cdot d \mathbf r = e \Phi,
\label{phase}
\end {equation}
where the integration ie performed along the closed path $\gamma = \gamma_1
\cup \gamma_2$, $\Phi$ being the total magnetic flux through the solenoid. The
condition for the absence of observable AB effect is $e\Phi = 2\pi n, \; n\in
\mathbb Z$.

Let us assume that the beam passes in the upper half of the space divided by
the plane $z=0$. Then the contribution of the string $S^\kappa_{\mathbf n}$ to
the relative phase shift of the wave function at the point $Q$ is found to be
\begin{align}
\Delta \varphi = 2\pi(1+\kappa)\mu \label{eq10}
\end{align}
and, if the beam passes in the lower half-space ($z < 0$), one has
\begin{align}
\Delta \varphi = 2\pi(\kappa -1)\mu \label{eq11}
\end{align}

It follows from the Eqs. (\ref{eq10}), (\ref{eq11}) the absence of AB effect
when $(1+\kappa)\mu$ and $(1-\kappa)\mu$, both are integers. In the case of
$\kappa \neq 0$, this yields immediately the following conditions: $2\mu \in
\mathbb Z$, that is the celebrated Dirac rule, and quantization of the string
weight, $2\kappa\mu \in \mathbb Z$. If $\kappa =0$, one obtains the Schwinger
quantization condition, $\mu \in \mathbb Z$.

At first sight the existence of the magnetic monopoles with an arbitrary
magnetic charge is in contradiction with the AB experiment. To clarify issue
let us recall that the relative phase shift (\ref{phase}) arises as result of
the parallel translation of wave function along the contour $\gamma$
surrounding the Dirac string. It is known that for the generators of
translations the Jacobi identity fails and for the finite translations one has
\cite{Jac,Gr}
\begin{align}
\bigl(U_{\mathbf a}U_{\mathbf b}\bigr)U_{\mathbf c} =\exp(i\alpha_3(\mathbf r
;\mathbf a, \mathbf b,\mathbf c)) U_{\mathbf a}\bigl(U_{\mathbf b}U_{\mathbf
c}\bigr) \label{as}
\end{align}
where $\alpha_s$ is the so-called {\em three cocycle}; $\alpha_3= 4\pi
\mu\,\mod 2\pi \mathbb Z$, if the monopole is enclosed by the simplex with
vertices $(\mathbf r,\mathbf r +\mathbf a, \mathbf r+ \mathbf a +\mathbf
b,\mathbf r+ \mathbf a+ \mathbf b +\mathbf c)$ and zero otherwise \cite{Jac}.
For the Dirac quantization condition being satisfied $\alpha_3 = 0 \mod 2\pi
\mathbb Z$, and (\ref{as}) provides an associative representation of the
translations, in spite of the fact that the Jacobi identity continues to fail.
Thus, we see that the AB effect requires more careful analysis, if one assumes
existence of an arbitrary monopole charge.

The emerging difficulties in explanation of the AB effect may be removed by
introducing nonassociative path-dependent wave function $\Psi(\mathbf r;
\gamma)$. Here we consider the simplified version of the approach developed in
\cite{N1,N2,N1a}.

Let us start with the nonassociative generalization of the gauge
transformations known as {\em gauge loop} $\mathcal Q$, which is related to the
so-called {\em string group} \cite{N1a,MJ}. The string group, denoted as
String$\,\mathfrak M$, is the group of all paths $\gamma$: $[0,1] \mapsto$
Diff$\,\mathfrak M$, where Diff$\,\mathfrak M$ denotes the diffeomorphism group
on $\mathfrak M = R^3\backslash\{0\}$ such that $ {\mathbf r} \mapsto {\mathbf
r}(t)= {\mathbf r}\gamma(t), \; t\in[0,1] $ and $\gamma(0)=$ identity; the
group composition is defined as $\gamma_{12}(t) = \gamma_1(t)\circ\gamma_2(t)$.

Now let $f_\gamma$ be the map:
\begin{equation}
 f_\gamma: \mathbf r
\mapsto U_{g_\gamma} = \exp{\big(i\alpha_1(\mathbf r;\gamma)\big)}\in \mathcal
Q,
\end{equation}
\begin{align}
\alpha_1(\mathbf r;\gamma)=  e \int_{\mathbf r}^{{\mathbf r}'} \mathbf
A(\boldsymbol \xi) \cdot d \boldsymbol \xi \label{g_1c},
\end{align}
where the integration is performed along a path connecting a point $\mathbf r$
with a point ${\mathbf r}' = \mathbf r\gamma(1)$, and $\mathbf A(\mathbf r)$ is
the vector potential.

The product of  two elements $U_{g_{\gamma_1}}, U_{g_{\gamma_2}} \in \mathcal
Q$ is defined as follows:
\begin{align}
&U_{g_{\gamma_1}}\ast U_{g_{\gamma_2}}=U_{g_{\gamma_1}{\mathbf\cdot}
{g_{\gamma_2}}}, \; U_{g_{\gamma_1}{\mathbf\cdot} {g_{\gamma_2}}} \in \mathcal
Q\\
&g_{\gamma_1}{\mathbf\cdot}{g_{\gamma_2}}=\alpha_1(\mathbf r; \gamma_1)
+\alpha_1({\mathbf r}_1; \gamma_2) +  \sigma (\mathcal C, \Sigma)), \nonumber
\end{align}
where ${\mathbf r}_1= \mathbf r\gamma_1(1)$, and $\sigma (\mathcal C, \Sigma)$
being the contribution of the Dirac string:
\begin{align}
\sigma = e\int_{\Sigma}{\mathbf h}\cdot d\mathbf S
\end{align}
is not zero if and only if the string $\mathcal C$ crosses the surface
$\Sigma$. The surface is parameterized as $\mathbf r(t,s)=\mathbf r\gamma_1(t)
\gamma_2(s)$ with  $0\leq s\leq t \leq 1$, and the vertices are $(\mathbf
r,{{\mathbf r}}_1,{{\mathbf r}}_2)$, where ${{\mathbf r}}_1 =\mathbf
r\gamma_1(1)$ and ${{\mathbf r}}_2 =\mathbf r\gamma_2(1)$.

For the triple-product we obtain the following result:
\begin{equation}
U_{g_{\gamma_1}}\ast (U_{g_{\gamma_2}}\ast U_{g_{\gamma_3}})= {\rm
e}^{i\alpha_3(\mathbf r; \gamma_1,\gamma_2,\gamma_3) }(U_{g_{\gamma_1}}\ast
U_{g_{\gamma_2}})\ast U_{g_{\gamma_3}} \nonumber
\end{equation}
where $\alpha_3 = 4\pi \mu$ is the three-cocycle. As is easy to see, if the
Dirac quantization condition is fulfilled, the operation $\ast$ is associative
and the local loop $\mathcal Q$ becomes the gauge group $U(1)$.

The nonassociative path dependent wave function is given by
\begin{equation}
\Psi(\mathbf r, \gamma)= {\rm e}^{i\alpha_1(\mathbf r, \gamma)} \Psi(\mathbf
r), \quad {\rm e}^{i\alpha_1(\mathbf r, \gamma)}\in {\rm QU(1)},
\end{equation}
and the realization of gauge loop $\mathcal Q$ in the space of the wave
functions $\Psi(\mathbf r, \gamma)$ is defined as follows:
\begin{equation}\label{A4}
U_{g_{\gamma_1}}\ast\Psi(\mathbf r; \gamma) ={\rm e}^{i(\alpha_1(\mathbf r;
\gamma_1) + \sigma (\mathcal C, \Sigma))}\Psi(\mathbf r' ; \gamma),
\end{equation}
where $U_{g_{\gamma_1}}=\exp({i\alpha_1(\mathbf r; \gamma_1)}\in{\mathcal Q }$,
and $\mathbf r'=\mathbf r\gamma(1)$.

For the product of $\Psi(\mathbf r, \gamma_1)$ and $\Psi(\mathbf r, \gamma_2)$
we obtain
\begin{equation}\label{A5}
\Psi(\mathbf r, \gamma_1)\ast\Psi(\mathbf r, \gamma_2) = {
e}^{i\alpha_2(\mathbf r; \gamma_1,\gamma_2)} \Psi(\mathbf r,\gamma_{12}),
\end{equation}
where
\begin{equation}
\alpha_{2}(\mathbf r;\gamma_{1},\gamma_{2})= e \int_{\Sigma} \mathbf B \cdot
d{\mathbf S} = e\Phi\big|_\Sigma \label{g_2c}
\end{equation}
$\Phi\big|_\Sigma$ being a magnetic flux through the two-dimensional simplex
$\Sigma$ with the vertices are $(\mathbf r,{{\mathbf r}}_1,{{\mathbf r}}_2)$,
where ${{\mathbf r}}_1 =\mathbf r\gamma_1(1)$ and ${{\mathbf r}}_2 =\mathbf
r\gamma_2(1)$. Finally, the triple-product obeys
\begin{align}\label{A6}
&\Psi(\mathbf r, \gamma_1)\ast \big(\Psi(\mathbf r, \gamma_2)\ast \Psi(\mathbf
r, \gamma_3)\big) \nonumber\\
& = {e}^{i\alpha_3(\mathbf r; \gamma_1,\gamma_2,\gamma_3) } \big(\Psi(\mathbf
r, \gamma_1)\ast \Psi(\mathbf r, \gamma_2)\big)\ast \Psi(\mathbf r, \gamma_3) .
\end{align}

Returning to the AB effect, let us consider the following paths
(Fig.\ref{solenoid}):
\begin{align}
\gamma_1(t): \mathbf r \rightarrow  \mathbf r + t(\mathbf a_1 + \mathbf a_2 ),\\
\gamma_2(t): \mathbf r \rightarrow  \mathbf r + t(\mathbf a_3 + \mathbf a_4 ).
\end{align}
Now, taking into account the formula (\ref{A5}) one can easily see that the
result of Eq.(\ref{AB}) is replaced by
\begin{align}
\Psi_Q& = \Psi_1(\mathbf r, \gamma_1) + \Psi_{2}(\mathbf r, \gamma_2) \nonumber\\
&= e^{i\alpha_1(\mathbf r, \gamma_1)}\psi_1(\mathbf r)+  e^{i\alpha_1(\mathbf
r, \gamma_2)}\psi_{2}(\mathbf r) \nonumber \\
&=  e^{i\alpha_1(\mathbf r, \gamma_2)}\big( { e}^{i\alpha_2(\mathbf r;
\gamma_1,\gamma_2)}\psi_1(\mathbf r) +\psi_{2}(\mathbf r)\big ), \label{AB_1}
\end{align}
and a relative phase shift $ \Delta \varphi$ is given by
\begin{equation}
\Delta \varphi = \alpha_{2}(\mathbf r;\gamma_{1},\gamma_{2})= e \int_{\Sigma}
\mathbf B \cdot d{\mathbf S} = e\Phi\big|_\Sigma \label{g_2d}
\end {equation}
where $\Phi\big|_\Sigma$ is {\em the magnetic flux of the monopole} through the
surface $\Sigma$ bounded by $\gamma_{1}$ and $\gamma_{2}$. Evidently, there is
{\em no any contribution from the Dirac string to the AB effect}.

Note, that for the true AB effect $\Phi\big|_\Sigma $ is equal to the total
magnetic flux$\Phi$ through the solenoid, and we obtain for the relative phase
shift $\Delta \varphi$ the same result as in (\ref{phase}),
\begin{equation}
\Delta \varphi = \int_{\Sigma} \mathbf B \cdot d{\mathbf S}= e\oint_\gamma
\mathbf A \cdot d \mathbf r = e \Phi.
\end {equation}

We close with some comments about relevance of this work to nonassociative
quantum mechanics. A standard quantum mechanics deals with a linear Hilbert
space and associative operators, therefore, the Dirac quantization rule is a
necessary condition for the self-consistence of quantum mechanics in the
presence of magnetic monopoles. Avoiding this condition implies introducing a
{\em nonassociative algebra of observables}. Since the nonassociativity
produces a serious conflict with a Hilbert space, one must define an
nonassociative equivalent to quantum mechanics, may be without Hilbert space
and in terms of density matrices alone \cite{Jac, Gr,G1,Gr1,G2}.

\section*{Acknowledgements}
The author is grateful to F.Aceves de la Cruz for helpful discussions.
 This work was partly supported by SEP-PROMEP (Grant No. 103.5/04/1911).

\bibliography{AB_loopf}
\end{document}